\documentstyle[pra,aps,multicol,psfig]{revtex}

\begin{document}
\draft

\title{Equivalence between Bell's inequality and a constraint \\
on stochastic field theories for EPR states}

\author{Lars M. Johansen}

\address{Department of Physics, University of Oslo, P.O.Box 1048
Blindern, N-0316 Oslo, Norway
\thanks{E-mail: Lars.M.Johansen@hibu.no}
\thanks{Permanent address: Buskerud College, P.O.Box
251, N-3601 Kongsberg, Norway}}

\date{\today}
\maketitle

\begin{abstract}

A generalized form of EPR state is defined, embracing both classical
and nonclassical states. It is shown that for such states, Bell's
inequality is equivalent to a constraint on stochastic field
theories. Thus, violation of Bell's inequality can be observed also
for weak violation of stochastic field theories. The Schr\"odinger
cat state is shown to be an example of this.

\end{abstract}

\pacs{03.65.Bz,42.50.Ar,42.50.Dv}

\begin{multicols}{2}

\section{Introduction}

In their classical paper \cite{Einstein35}, Einstein, Podolsky and
Rosen (EPR) considered a twoparticle state. They pointed out that for
a suitable choice of experimental parameters, it was possible to
``predict with certainty" the outcome of a measurement on particle 2
on basis of a measurement on particle 1. They claimed that ``there
exists an element of physical reality corresponding to this physical
quantity". Since these elements of reality are not reflected in the
theory, they concluded that quantum theory is incomplete. They saw
that alternatively the reality of the second system might ``depend
upon the process of measurement carried out on the first system,
which does not disturb the second system in any way". This is the
possibility of nonlocality, which they immediately rejected: ``No
reasonable definition of reality could be expected to permit this."

EPR considered position and momentum observables, which have a
continuous spectrum. Bohm reformulated the EPR problem in terms of
dichotomic observables by using the singlet spin-$\frac{1}{2}$ state
\cite{Bohm51}. Also for this state it is possible to ``predict with
certainty" the outcome of a spin measurement on particle 2 on basis
of a measurement on particle 1. A particular orientation of the
spin-filters must be used to achieve this.

Bell derived a contradiction with quantum theory by using EPR's two
fundamental assumptions of locality and realism \cite{Bell64}. He
employed the singlet spin-$\frac{1}{2}$ state introduced by Bohm.

Greenberger, Horne and Zeilinger found a contradiction between
quantum theory and local realism without the use of inequalities
\cite{Greenberger89}. They employed a three-particle entangled state
where it was possible to ``predict with certainty" the outcome of a
measurement at two different locations on basis of a single
measurement in a third location.

Hardy derived contradictions between quantum theory and local realism
without the use of inequalities for two-particle states
\cite{Hardy92,Hardy92b}. Also here, ``elements of reality" played
an essential role in the derivation.

Gisin and Peres recently showed that any nonfactorable pure state
violates Bell's inequality \cite{Gisin92}. A similar result was
derived by Mann, Revzen and Schleich \cite{Mann92}. These proofs
employ operators for which no general measurement method is known.
The experimental testing of local realism has only been performed
with entangled two-particle states (reviews of experiments are given
in Refs. \cite{Clauser78,Ballentine87,Home91,Chiao95}). There does
not exist any general experiment which can be used to test whether an
arbitrary state violates local realism.

In this paper, I consider in particular a family of states which will
be called EPR states. These states allow a property of one subsystem
to be predicted with certainty from a measurement on another
subsystem with spacelike separation. According to the definition
given here, an EPR state may be either classical or nonclassical, it
may be a multiparticle state or possess an indefinite particle
number, and it may be either pure or mixed. Several examples of EPR
states are given, such as entangled  states, single photon states,
coherent states and ``Schr\"odinger cat" states. In fact, {\em any}
single mode state, pure or mixed, can be transformed into an EPR
state by use of beamsplitters.

I demonstrate that any EPR state which violates an inequality for
stochastic field theories also violates Bell's inequality, and vice
versa.

In stochastic field theories the Glauber-Sudarshan $P$-distribution
is nonnegative and not more singular than a delta function
\cite{Mandel95b}. Some quantum states violate such inequalities
(reviews can be found in Refs. \cite{Loudon80,Reid86}). Other
researchers have found that strong violation of a Cauchy-Schwarz
inequality and stochastic field theories is required to observe
violation of Bell's inequality \cite{Reid86,Su91}. I show that
violation of local realism can also be observed for weak violation of
stochastic field theories.

\section{Elements of reality and correlation strength}

In this section, I propose a definition for a generalized form of EPR
states. A normalized correlation is introduced, and it is shown that
the correlation strength is maximized in an EPR state. The
correlation strength is expressed in terms of operators for the
input channels to the interferometer.

\subsection{A general experiment}

Consider the experiment depicted in Fig. \ref{fig:general}. The
lefthand and righthand side of the experiment is designated by
indices $k=1$ and 2, respectively. A phase delay $\theta_k$ is
inserted into channel $b_k$. Afterwards this channel is mixed with
channel $a_k$ on a semireflecting beamsplitter $BS_k$, yielding
output channels $c_k$ and $d_k$.

Let the photon number operator for channel $\mu_k$ be designated by
$\hat{n}_{\mu_k}$ ($\mu=a,b,c,d$ and $k=1,2$). We shall say that a
state is an {\em EPR state} if it is possible to find a parameter
choice ($\theta_1,\theta_2$) so that
\begin{mathletters}
\begin{eqnarray}
	\langle \hat{n}_{c_1} \hat{n}_{d_2} \rangle =
	\langle \hat{n}_{d_1} \hat{n}_{c_2} \rangle &=& 0,
	\label{eq:anticorr1} \\
	\langle \hat{n}_{c_1} \hat{n}_{c_2} \rangle +
	\langle \hat{n}_{d_1} \hat{n}_{d_2} \rangle &\ne& 0.
	\label{eq:coinc1}
\end{eqnarray}
	\label{eq:element1}
\end{mathletters}
If $\langle \hat{n}_{\mu_1} \hat{n}_{\nu_2} \rangle=0$
($\mu,\nu=c,d$), the joint probability of finding at least one photon
both in channels $\mu_1$ and $\nu_2$ vanishes. On the other hand, if
$\langle \hat{n}_{\mu_1} \hat{n}_{\nu_2} \rangle > 0$, this same
probability is nonvanishing. Still, in the latter case, there may of
course be a nonvanishing probability of finding zero photons in at
least one of the two channels. On basis of these considerations, we
conclude that that a state satisfying the conditions
(\ref{eq:element1}) support EPR elements of reality in the following
sense: If at least one photon is found in channel $c_1$ ($d_1$), it
can be predicted with certainty that
\begin{itemize}
\item no photons will be found in channel $d_2$ ($c_2$)
\item zero or more photons will be found in channel $c_2$ ($d_2$).
\end{itemize}
The maximally entangled state which is usually employed in Bell-type
experiments is an EPR state according to the definition above.
However, this definition even encompasses states where for a certain
subsensemble no coincidences occur between the two sides of the
interferometer. Moreover, we shall see that not all EPR states
violate local realism. In fact, EPR states can be generated from {\em
any} single-mode state. Several example states illustrating these
effects will be considered in section \ref{sec:examples}.

We introduce operators for the photon number difference and the
photon number sum for the output channels from beamsplitter $BS_k$,
\begin{mathletters}
\begin{eqnarray}
	\hat{D}_k &=& \hat{n}_{c_k} - \hat{n}_{d_k}, \\
	\hat{S}_k &=& \hat{n}_{c_k} + \hat{n}_{d_k}.
\end{eqnarray}
	\label{eq:sumdiff}
\end{mathletters}
The correlation between the leftside and rightside interferometers
can be quantified by the normalized ratio
\begin{equation}
	E = {\langle \hat{D}_1 \hat{D}_2 \rangle \over \langle \hat{S}_1
	\hat{S}_2 \rangle}.
\end{equation}
Since the photon number difference cannot exceed the photon number
sum, the modulus of $E$ is restricted to unity,
\begin{equation}
	\mid E \mid \le 1.
	\label{eq:Elimit}
\end{equation}
Using the definitions (\ref{eq:sumdiff}), we may write
\begin{equation}
	E = {\langle \hat{n}_{c_1}
	\hat{n}_{c_2} \rangle - \langle \hat{n}_{c_1} \hat{n}_{d_2}
	\rangle - \langle \hat{n}_{d_1} \hat{n}_{c_2} \rangle + \langle
	\hat{n}_{d_1} \hat{n}_{d_2} \rangle \over \langle \hat{n}_{c_1}
	\hat{n}_{c_2} \rangle + \langle \hat{n}_{c_1} \hat{n}_{d_2}
	\rangle + \langle \hat{n}_{d_1} \hat{n}_{c_2} \rangle + \langle
	\hat{n}_{d_1} \hat{n}_{d_2} \rangle}.
\end{equation}
If the conditions (\ref{eq:element1}) are fulfilled, it follows that
$E = 1$.

Assume that the annihilation and creation operators for channel
$\mu_k$ are designated by $\hat{\mu}_k$ and $\hat{\mu}_k^{\dag}$,
respectively. The annihilation operators for the input and output
channels are connected by a unitary transformation
\begin{equation}
 \left ( \begin{array}{c} \hat{c}_k \\ \hat{d}_k \end{array} \right )
 = {1 \over \sqrt{2}} \left ( \begin{array}{cc} 1 & 1 \\ -1 & 1
 \end{array} \right ) \left ( \begin{array}{cc} 1 & 0 \\ 0 & e^{i
 \theta_k} \end{array} \right ) \left ( \begin{array}{c}  \hat{a}_k
 \\ \hat{b}_k \end{array} \right ).
\end{equation}
Using this, the operators for the photon number difference and sum
may be written as
\begin{mathletters}
\begin{eqnarray}
	\hat{D}_k &=& \hat{a}_k^{\dag} \hat{b}_k e^{i \theta_k} +
	\hat{a}_k \hat{b}_k^{\dag} e^{- i \theta_k}, \\
	\hat{S}_k &=& \hat{n}_{a_k} + \hat{n}_{b_k}.
\end{eqnarray}
\end{mathletters}
Note, in particular, that only the photon number difference is
modified by phase changes. It can be shown \cite{Tan90} that $E$ may
be written in the form
\begin{equation}
 E = A_1 \cos(\theta_1 - \theta_2 + \xi) + A_2 \cos (\theta_1 +
 \theta_2 + \zeta),
 \label{eq:trigform}
\end{equation}
The coefficients $A_k$ will be called ``correlation-amplitudes". They
are nonnegative, and defined as
\begin{mathletters}
\begin{eqnarray}
  A_1 &=& {2 \mid \langle \hat{a}_1^{\dag} \hat{b}_1 \hat{a}_2
  \hat{b}_2^{\dag} \rangle \mid \over \langle ( \hat{n}_{a_1} +
  \hat{n}_{b_1}) ( \hat{n}_{a_2} +  \hat{n}_{b_2}) \rangle}, \\
  A_2 &=& {2 \mid \langle \hat{a}_1^{\dag} \hat{b}_1 \hat{a}_2^{\dag}
  \hat{b}_2 \rangle \mid \over \langle ( \hat{n}_{a_1} +
  \hat{n}_{b_1}) ( \hat{n}_{a_2} +  \hat{n}_{b_2}) \rangle},
\end{eqnarray}
	\label{eq:amplitudes}
\end{mathletters}
where
\begin{mathletters}
\begin{eqnarray}
  \xi = \arg\langle \hat{a}_1^{\dag} \hat{b}_1 \hat{a}
  \hat{b}_2^{\dag} \rangle,\\
  \zeta = \arg \langle \hat{a}_1^{\dag} \hat{b}_1 \hat{a}_2^{\dag}
  \hat{b}_2 \rangle.
\end{eqnarray}
\end{mathletters}
According to Eq. (\ref{eq:trigform}), $E$ can be modulated between
$\pm A$, where
\begin{equation}
	A = A_1 + A_2.
	\label{eq:totalamplitude}
\end{equation}
This quantity can be regarded as a normalized measure of the
correlation strength. Also, it can be seen as the total correlation
amplitude. It is always nonnegative, and cannot exceed unity since
the modulus of $E$ cannot exceed unity. For EPR states
(\ref{eq:element1}) we have
\begin{equation}
	A=1.
\end{equation}
Conversely, if $A=1$, it can be shown that the state supports EPR
elements or reality in the sense of Eq. (\ref{eq:element1}).
Moreover, if $A=1$, a parameter choice can be found for which $E=-1$,
and for which
\begin{mathletters}
\begin{eqnarray}
	\langle \hat{n}_{c_1} \hat{n}_{c_2} \rangle =
	\langle \hat{n}_{d_1} \hat{n}_{d_2} \rangle &=& 0,
	\\
	\langle \hat{n}_{c_1} \hat{n}_{d_2} \rangle +
	\langle \hat{n}_{d_1} \hat{n}_{c_2} \rangle &\ne& 0.
\end{eqnarray}
\end{mathletters}

\subsection{Homodyne detection}

In the special case where the channels $b_k$ are coherent state local
oscillators (see Fig. \ref{fig:homodyne}), the density operator may
be written as
\begin{equation}
	\hat{\rho} = \hat{\rho}_a \otimes \hat{\rho}_b,
\end{equation}
where
\begin{equation}
	\hat{\rho}_b = \hat{\rho}_{b_1} \otimes \hat{\rho}_{b_2}
\end{equation}
and
\begin{equation}
	\hat{\rho}_{b_k} = \mid \beta_k e^{i \theta_k} \rangle_{b_k}
	\quad _{b_k}{\langle \beta_k e^{i \theta_k} \mid}.
\end{equation}
Now the local oscillator phases $\theta_k$ play the role of local
parameters. The amplitudes $\beta_k$ may be chosen to be real, and
it can be shown \cite{Tan90,Johansen96} that $E$ is maximized by the
choice
\begin{mathletters}
\begin{eqnarray}
 \beta_1 \beta_2 &=& \sqrt{\langle  \hat{n}_{a_1} \hat{n}_{a_2}
 \rangle}, \\ {\beta_1 \over \beta_2} &=& \sqrt{{\langle
 \hat{n}_{a_1} \rangle \over \langle \hat{n}_{a_2} \rangle }}.
\end{eqnarray}
 \label{eq:optimalamplitudes}
\end{mathletters}
This leads to the following form \cite{Johansen96} for the
correlation amplitudes $A_k$,
\begin{mathletters}
\begin{eqnarray}
	A_1 = { \mid g^{(1,1)} \mid \over 1 + \sqrt{g^{(2,2)}}},\\
	A_2 = { \mid g^{(2,0)} \mid \over 1 + \sqrt{g^{(2,2)}}},
\end{eqnarray}
	\label{eq:homodynecoeff}
\end{mathletters}
where
\begin{mathletters}
\begin{eqnarray}
    g^{(1,1)} &=& {\langle \hat{a}_1^{\dag} \hat{a}_2
    \rangle \over \sqrt{\langle \hat{n}_{a_1} \rangle
    \langle \hat{n}_{a_2} \rangle}},\\
    g^{(2,0)} &=& {\langle \hat{a}_1^{\dag} \hat{a}_2^{\dag}
    \rangle \over \sqrt{\langle \hat{n}_{a_1} \rangle
    \langle \hat{n}_{a_2} \rangle}},\\
    g^{(2,2)} &=& {\langle \hat{a}_1^{\dag} \hat{a}_2^{\dag}
    \hat{a}_2 \hat{a}_1 \rangle \over \langle \hat{n}_{a_1} \rangle
    \langle \hat{n}_{a_2} \rangle}.
\end{eqnarray}
\end{mathletters}
We see that the correlation amplitudes $A_k$ can be defined in terms
of Glauber coherence functions \cite{Glauber63a}. In particular,
$g^{(1,1)}$ and $g^{(2,2)}$ are known as the degree of first and
second order coherence (also sometimes called the degree of second
and fourth order coherence). They can be observed, e.g., as the
interference visibility and the coincidence rate in a Mach-Zehnder
interferometer.

\section{Correlation inequalities}

In this section, inequalities are derived for the correlation
amplitudes $A_k$ in stochastic field theories, locally realistic
theories and quantum theory. The connection between the inequalities
is discussed, particularly for EPR states.

\subsection{An inequality for stochastic field theories}

In appendix \ref{sec:app}, it is shown that stochastic field theories
impose the restrictions
\begin{equation}
	A_k \le {1 \over 2} \quad (k=1,2).
	\label{eq:stochastic}
\end{equation}
Thus stochastic field theories restrict each correlation amplitude in
itself. Both amplitudes may reach the maximal value of $1/2$
simultaneously. This is seen, e.g., in a coherent state.

\subsection{Bell's inequality}

It has been shown that in local, realistic theories, the quantity
\begin{equation}
 B = E(\theta_1,\theta_2) - E(\theta_1',\theta_2) +
 E(\theta_1,\theta_2') + E(\theta_1',\theta_2')
 \label{eq:CHSH}
\end{equation}
is restricted by the condition \cite{Clauser78,Reid86}
\begin{equation}
	\mid B \mid \le 2.
\end{equation}
Tan {\em et al.} \cite{Tan90} showed that by a proper choice of
phases $\theta_k$, the maximal value of $B$ is
\begin{equation}
	B_{\max} = 2 \sqrt{2} \: \sqrt{A_1^2+A_2^2}.
	\label{eq:CHSHmax}
\end{equation}
It follows that local realism is violated unless
\begin{equation}
	A_1^2 + A_2^2 \le {1 \over 2}.
	\label{eq:local}
\end{equation}
This is a necessary but not sufficient condition for a state to be
describable in terms of a local, hidden variable theory.

\subsection{Quantum inequalities}

If no interconnection existed between the four terms in the
expression $B$ in Eq. (\ref{eq:CHSH}), it would have a maximum of 4.
However, we saw that in locally realistic theories the maximum is 2,
and it follows from the considerations above that the same limit
applies in stochastic field theories. Tsirelson showed that in
quantum theory the allowed maximum is \cite{Cirelson80}
\begin{equation}
	\mid B \mid \le 2 \sqrt{2}.
\end{equation}
It follows from (\ref{eq:CHSHmax}) that Tsirelson's inequality can be
written as
\begin{equation}
	A_1^2 + A_2^2 \le 1.
	\label{eq:tsirelson}
\end{equation}
This is a necessary, but not a sufficient condition on the amplitudes
$A_k$. A necessary and sufficient condition within quantum theory is
found by combining Eqs. (\ref{eq:Elimit}) and
(\ref{eq:totalamplitude}),
\begin{equation}
	A_1 + A_2 \le 1.
	\label{eq:quantumlimit}
\end{equation}

\subsection{Comparison of the inequalities}

\begin{figure}
\centerline{\psfig{figure=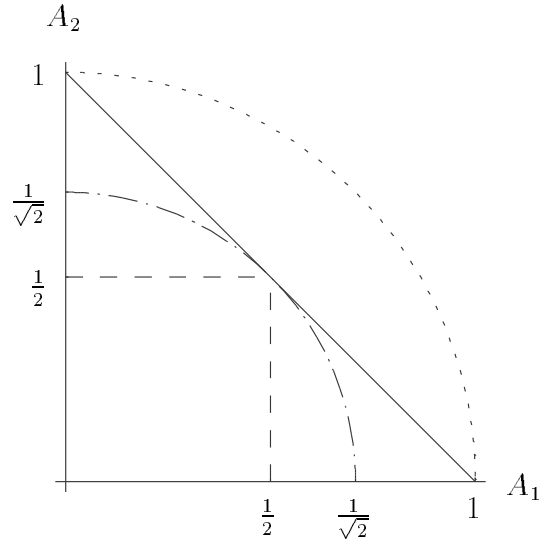}}
\caption{The limits on the correlation amplitudes $A_k$. Solid line:
Constraint on quantum theory. This line also defines EPR states.
Dashed line: Constraint on stochastic field theories. Dash-dotted
line: Bell's inequality. Dotted line: Tsirelson's inequality.}
\label{fig:limits}
\end{figure}

The inequalities (\ref{eq:stochastic}), (\ref{eq:local}),
(\ref{eq:tsirelson}) and (\ref{eq:quantumlimit}) have been
illustrated in Fig. \ref{fig:limits}. We see that stochastic
field theories allow the smallest range of amplitudes $A_k$. Within
locally realistic theories, a larger range of amplitudes is allowed,
and an even larger range of amplitudes is permitted in quantum
theory. The widest range of amplitudes is allowed by the Tsirelson
inequality, but it is seen to permit amplitudes forbidden by quantum
theory.

It is interesting to note that the limits imposed by stochastic field
theories, local realism and quantum theory intersect in the point
$A_1=A_2=1/2$. Note furthermore that EPR states are represented by
the limit line for quantum theory. Thus we see that any EPR state for
which $A_k \ne 1/2$ violates both local realism and stochastic field
theories.The farther away from the central point $A_1=A_2=1/2$, the
stronger the violation.

If we consider states where one amplitude $A_k$ is zero, the maximal
amplitude allowed by stochastic field theories is $1/2$, in locally
realistic theories it is $1/\sqrt{2}$ and in quantum theory it is 1.
This is somewhat reminiscent of the result found by Su and
W\'odkiewicz \cite{Su91}. They examined the interference visibility
of the intensity correlation in two-photon experiments, and found
that stochastic field theories restricts this visibility to $1/2$,
local realism to $1/\sqrt{2}$ and quantum theory to 1. Their
conclusion was that violation of local realism requires strong
violation of classical field theory. This is in agreement with the
results found here, provided that one amplitude $A_k$ vanishes. For
an entangled two-photon state, one of the amplitudes $A_k$ must
vanish (see Sec. \ref{sec:examples}). However, Fig. \ref{fig:limits}
shows that in EPR states, violation of local realism can be observed
also for {\em weak} violation of stochastic field theories.

Another interesting observation is that if one of the amplitudes
$A_k$ exceeds the limit of $1/2$ imposed by stochastic field
theories, then according to the quantum limit (\ref{eq:quantumlimit})
the other amplitude must be smaller than $1/2$. Thus, although
quantum theory allows the amplitudes to exceed the classical limit
(\ref{eq:stochastic}), it instead imposes a complementarity relation
(\ref{eq:quantumlimit}) between the two.

\section{Some EPR states}
\label{sec:examples}

In this section, various EPR states are considered. They are mostly
well known, some classical and some nonclassical.

\begin{figure}
\centerline{\psfig{figure=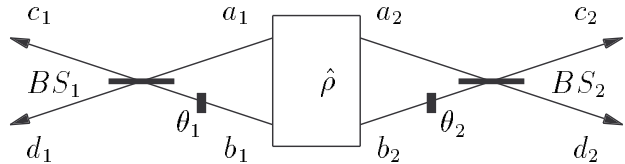}}
\caption{A general Bell experiment.}
\label{fig:general}
\end{figure}

\begin{figure}
\centerline{\psfig{figure=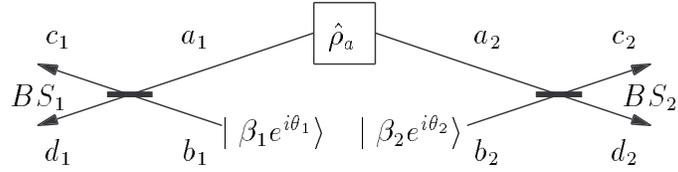}}
\caption{A Bell experiment involving two local oscillators.}
\label{fig:homodyne}
\end{figure}

The section is divided into two subsections for two different
experimental setups; a general setup (Fig. \ref{fig:general}) and a
setup using coherent local oscillators (Fig. \ref{fig:homodyne}).
The task is essentially to describe possible contents in the ``black
boxes" in Figs. \ref{fig:general} and \ref{fig:homodyne}.

\subsection{The general setup}

\subsubsection{An arbitrary single mode state}

Consider an arbitrary single mode state, pure or mixed. Assume that
this state is first mixed with vacuum on a semireflecting
beamsplitter (see Fig. \ref{fig:gensplit}). Next, assume that each
output channel from this beamsplitter is again mixed with vacuum on a
semireflecting beamsplitter. It can then be shown (see App.
\ref{sec:gensplit}) that this produces an EPR state with the
properties $A_1 = A_2 = \frac{1}{2}$. Such states therefore are
classical in the sense that they do not violate the Cauchy-Schwarz
and
Bell inequalities (\ref{eq:stochastic}) and (\ref{eq:local}).

It is interesting that a highly coherent state can be produced from
any single mode state. This shows that a quantized field behaves in
many ways just like a classical field.

\subsubsection{Entangled states}

A maximally entangled state can be written as \cite{Horne89}
\begin{eqnarray}
	\mid \psi \rangle = {1 \over \sqrt{2}} ( &\mid& 1
	\rangle_{a_1} \otimes \mid 0 \rangle_{b_1} \otimes \mid 1
	\rangle_{a_2} \otimes \mid 0 \rangle_{b_2} \nonumber \\ + &\mid&
	0 \rangle_{a_1} \otimes \mid 1 \rangle_{b_1} \otimes \mid 0
	\rangle_{a_2} \otimes \mid 1 \rangle_{b_2} ).
\end{eqnarray}
The photons are either in channel $a_1$ and $a_2$ or in channels
$b_1$ and $b_2$. This is equivalent to the situation in the singlet
spin-$\frac{1}{2}$ state, where either the left spin is up and the
right is down or vice versa. For this state, $A_1=0$ and $A_2=1$.
Thus the state is both an EPR state and it yields maximal violation
of local realism. An equivalent maximally entangled state is
\begin{eqnarray}
	\mid \psi \rangle = {1 \over \sqrt{2}} ( &\mid& 1
	\rangle_{a_1} \otimes \mid 0 \rangle_{b_1} \otimes \mid 0
	\rangle_{a_2} \otimes \mid 1 \rangle_{b_2} \nonumber \\ + &\mid&
	0 \rangle_{a_1} \otimes \mid 1 \rangle_{b_1} \otimes \mid 1
	\rangle_{a_2} \otimes \mid 0 \rangle_{b_2} ).
\end{eqnarray}
Here the photons are either in channel $a_1$ and $b_2$ or $b_1$ and
$a_2$. For this state, $A_1=1$ and $A_2=0$.

\subsubsection{Two independent photons}

\begin{figure}
\centerline{\psfig{figure=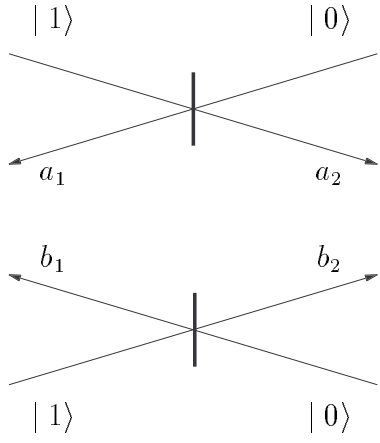}}
\caption{Generating an EPR state from two independent photons}
\label{fig:independ}
\end{figure}

Consider the two-photon state
\begin{equation}
	\mid \psi \rangle = \mid 1 \rangle \: \otimes \mid 1 \rangle.
\end{equation}
Such states can be generated, e.g., in parametric down-conversion.
Due to the product form, the two photons are independent. Assume that
each photon is mixed with vacuum on a beamsplitter (cf. Fig.
\ref{fig:independ}). In terms of output states from the beamsplitter,
this may be written
\begin{eqnarray}
	\mid \psi \rangle = \frac{1}{2} ( \mid 1 \rangle_{a_1}
	\otimes \mid 0 \rangle_{a_2} + \mid 0 \rangle_{a_1} \otimes \mid
	1 \rangle_{a_2} &)& \nonumber \\ \otimes ( \mid 1 \rangle_{b_1}
	\otimes \mid 0 \rangle_{b_2}  + \mid 0 \rangle_{b_1} \otimes \mid
	1 \rangle_{b_2} &)&.
\end{eqnarray}
This is still a product state between the $a$- and $b$-channels.
There has been some discussion whether such states can violate local
realism, the argument being that it is really generated from a
product state \cite{DeCaro94,Kwiat95}. However, the $a$-channels
are later pairwise mixed with $b$-channels, and there is no longer a
product form between eigenstates for the left and right sides of the
interferometer. Here it is found that $A_1 = 1$, $A_2=0$. Thus this
is an EPR state which violates local realism maximally. Note that EPR
elements of reality can only be predicted in 50\% of the outcomes,
because in the rest 50\% of the outcomes, the two photons will go to
the same side of the interferometer.

\subsection{A setup with local oscillators}

\subsubsection{Coherent states}

For coherent states
\begin{equation}
	\mid \psi \rangle_{a} = \mid \alpha_1 \rangle_{a_1} \otimes \mid
	\alpha_2 \rangle_{a_2}
\end{equation}
we find that $\mid g^{(1,1)} \mid = \mid g^{(2,0)} \mid = g^{(2,2)} =
1$. It therefore  follows from Eqs. (\ref{eq:homodynecoeff}) that
$A_k=1/2$. Thus, coherent states are EPR states, and they neither
violate Bell's inequality (\ref{eq:local}) nor the constraint
(\ref{eq:stochastic}) on stochastic field theories. This is in
agreement with the conclusions in Ref. \cite{Mann92}.

\subsubsection{A split single photon}

Consider the split single photon state
\cite{Tan90,Oliver89,Tan91,Hardy94}
\begin{equation}
	\mid \psi \rangle_a = {1 \over \sqrt{2}} \left ( \mid 1
	\rangle_{a_1} \otimes \mid 0 \rangle_{a_2} + \mid 0 \rangle_{a_1}
	\rangle_{b_1} \otimes \mid 1 \rangle_{a_2} \right ).
\end{equation}
It yields $g^{(1,1)}=1$, $g^{(2,0)}=0$, $g^{(2,2)}=0$, and thus
$A_1=1$, $A_2=0$. Thus it yields a strong violation both of local
realism and of stochastic field thories.

Note that according to the conditions (\ref{eq:optimalamplitudes}),
the local oscillator amplitudes should vanish. This is not possible
in practice if the purpose is to measure $E$, since then no
coincidences occur at all. However, for sufficiently small local
oscillator amplitudes, the state behaves ``almost" as an EPR state
\cite{Tan91}.

\subsubsection{A split ``Schr\"odinger cat"}
\label{sec:cat}

As an example of an EPR state yielding both weak violation of
Bell's inequality and the constraint (\ref{eq:stochastic}) on
stochastic field theories, consider the state
\begin{equation}
	\mid \psi \rangle_a = N \, \left ( \mid \alpha \rangle_{a_1} \:
	\otimes \mid \alpha \rangle_{a_2} + e^{i \phi} \mid -\alpha
	\rangle_{a_1} \: \otimes \mid -\alpha \rangle_{a_2} \right ),
\end{equation}
where the normalization constant is
\begin{equation}
	N = \left [ \, 2 \left (1  +  e^{-4 \mid \alpha \mid ^2 \,} \cos
	\phi  \right ) \right ]^{-1/2}.
\end{equation}
These states can be generated from a ``Schr\"odinger cat state"
\cite{Yurke86}
\begin{equation}
	\mid \psi \rangle = N \, \left ( \, \mid \sqrt{2} \alpha
	\rangle
	+ e^{i \phi} \mid - \sqrt{2} \alpha \rangle \, \right )
\end{equation}
by mixing with vacuum at a semireflecting beamsplitter. For parameter
choices $\phi=0$ and $\phi=\pi$ we have even and odd coherent states
\cite{Dodonov74}, while $\phi=\pi/2$ yield ``Yurke-Stoler" states
\cite{Yurke86}. Atomic Schr\"odinger cat states have recently been
generated experimentally \cite{Monroe96,Noel96}. It can be shown that
\begin{mathletters}
\begin{eqnarray}
	g^{(1,1)} &=& 1, \\
	g^{(2,0)} &=&  {1 + e^{-4 \mid \alpha \mid^2} \cos \phi \over
	1 - e^{-4 \mid \alpha \mid^2} \cos \phi}, \\
	g^{(2,2)} &=& \left ( g^{(2,0)} \right )^2.
\end{eqnarray}
\end{mathletters}
By inserting into Eqs. (\ref{eq:homodynecoeff}), it follows that
\begin{mathletters}
\begin{eqnarray}
	A_1 &=& {1 \over 2} \left ( 1 - e^{-4 \mid \alpha \mid^2} \cos
	\phi \right ), \\
	A_2 &=& {1 \over 2} \left ( 1 + e^{-4 \mid \alpha \mid^2} \cos
	\phi \right ).
\end{eqnarray}
\end{mathletters}
It is easily seen that
\begin{equation}
	A_1 + A_2 = 1.
\end{equation}
The Schr\"odinger cat states are therefore EPR states, regardless of
the phase choice $\phi$. It also follows that
\begin{equation}
	A_1^2 + A_2^2 = {1 \over 2} \left [ 1 + e^{-8 \mid \alpha \mid^2}
	\cos^2 \phi \right ].
\end{equation}
Thus, both Bell's inequality and inequality (\ref{eq:stochastic}) are
violated for any choice of parameters except when $\phi=\pi/2$. Even
if the parameters are chosen so that only a weak violation of
stochastic field theories and inequality (\ref{eq:stochastic}) takes
place, Bell's inequality is also violated. Note, however, that
although these states display a small violation of local realism even
for macroscopic amplitudes $\mid \alpha \mid$, this violation
vanishes exponentially. Therefore, these states are not suitable for
demonstrating violation of local realism in macroscopic states.

\section{Conclusion}

It has been shown that violation of local realism can be observed
also for weak violation of classical field theories. It was shown
that Bell's inequality is equivalent to a constraint on classical
field theories.

Violation of local realism requires that a certain two-point
correlation is stronger than classically \cite{Peres78}. However, it
also requires that another correlation form is reduced {\em below}
the classical limit, as was shown in this paper.

The Bell inequality used in this paper involves an additional
assumption related to the ``no-enhancement assumption"
\cite{Clauser78,Reid86,Santos92,Tan92}. Therefore, the experiments
discussed in this paper may rule out only hidden variable theories
which fulfill this assumption. This is a common feature of all Bell
inequalities that have been experimentally tested so far.
Inequalities derived without this assumption in general require
higher detector efficiencies in order to be tested
\cite{Clauser78,Garg87}.

There exist EPR states which display nonclassical properties but
which nevertheless do not violate the constraint
(\ref{eq:stochastic}) on classical theories. An example of this is
the split Yurke-Stoler state (see section \ref{sec:cat}). Such states
of course do not violate the Bell-inequality (\ref{eq:local}) either.
Thus, the experiment presented here does not demonstrate violation of
local realism of every nonclassical EPR state. However, if the EPR
state violates the classical inequality (\ref{eq:stochastic}), it
also violates the Bell inequality (\ref{eq:local}).

It should be noted that whereas a classical Glauber-Sudarshan $P$-
distribution is a sufficient condition for a state to obey local
realism, a nonclassical $P$-distribution is in general only a
necessary and not a sufficient condition for the violation of local
realism. One may imagine, e.g., a product state where either of the
sub-states possess some highly nonclassical $P$-distribution. Such a
state does not violate any Bell inequality \cite{Gisin92}.

\section*{Acknowledgments}

This research was financed by the University of Oslo, and is a
cooperative project with Buskerud College.

\end{multicols}

\appendix

\section{Constraint on stochastic field theories}
\label{sec:app}

The density operator for the experiment shown in Fig.
\ref{fig:general} may be represented in terms of a Glauber-Sudarshan
quasi-distribution $P(\alpha_1,\alpha_2,\beta_1,\beta_2)$ as
\begin{eqnarray}
    \hat{\rho} &=& \int P(\alpha_1,\alpha_2,\beta_1,\beta_2)
	\mid \alpha_1 \rangle_{a_1} \: _{a_1}{\langle \alpha_1 \mid}
	\: \otimes
	\mid \alpha_2 \rangle_{a_2} \: _{a_2}{\langle \alpha_2 \mid}
	\nonumber \\ &\otimes&
    \mid \beta_1 \rangle_{b_1} \: _{b_1}{\langle \beta_1 \mid}
	\: \otimes
    \mid \beta_2 \rangle_{b_2} \: _{b_2}{\langle \beta_2 \mid}
	\: d^2 \alpha_1 \: d^2 \alpha_2 \: d^2 \beta_1 \: d^2 \beta_2.
\end{eqnarray}
According to Eqs. (\ref{eq:amplitudes}), the amplitudes $A_k$ may be
written as
\begin{mathletters}
\begin{eqnarray}
  A_1 &=& {2 \mid \int P(\alpha_1,\alpha_2,\beta_1,\beta_2)
  \alpha_1^* \beta_1 \alpha_2^* \beta_2 \: d^2 \alpha_1 \: d^2
  \alpha_2 \: d^2 \beta_1 \: d^2 \beta_2 \mid \over \int P(\alpha_1,
  \alpha_2, \beta_1,  \beta_2) ( \mid \alpha_1 \mid^2 + \mid \beta_1
  \mid^2)( \mid  \alpha_2 \mid^2 + \mid \beta_2 \mid^2) \: d^2
  \alpha_1 \: d^2 \alpha_2 \: d^2 \beta_1 \: d^2 \beta_2 } , \\
  A_2 &=& {2 \mid \int P(\alpha_1, \alpha_2, \beta_1, \beta_2)
  \alpha_1^* \beta_1^* \alpha_2 \beta_2 \: d^2 \alpha_1 \: d^2
  \alpha_2 \: d^2 \beta_1 \: d^2 \beta_2 \mid \over \int P(\alpha_1,
  \alpha_2, \beta_1, \beta_2) ( \mid \alpha_1 \mid^2 + \mid \beta_1
  \mid^2)( \mid \alpha_2 \mid^2 + \mid \beta_2 \mid^2) \: d^2
  \alpha_1 \: d^2 \alpha_2 \: d^2 \beta_1 \: d^2 \beta_2 }.
\end{eqnarray}
\end{mathletters}
For any complex numbers $\alpha_k$ and $\beta_k$ the inequality
\begin{equation}
	\mid \alpha_k \mid^2 + \mid \beta_k \mid^2 \: \ge \: 2 \mid
	\alpha_k \beta_k \mid
\end{equation}
applies. This inequality was recently used to demonstrate a
nonclassical two-photon effect \cite{Franson91,Franson91b}. If the
state can be described in terms of stochastic field theories, the
$P$-distribution is nonnegative and not more singular than a
delta-function \cite{Mandel95b}. For such $P$-distributions
it follows that \begin{mathletters}
\begin{eqnarray}
	A_1 \: &\le& \: {\mid \int P(\alpha_1,\alpha_2,\beta_1,\beta_2)
	\alpha_1^*\beta_1 \alpha_2^* \beta_2 \: d^2 \alpha_1 \: d^2
	\alpha_2 \: d^2 \beta_1 \: d^2 \beta_2 \mid \over 2
	\int P (\alpha_1, \alpha_2, \beta_1, \beta_2) 4 \mid \alpha_1
	\alpha_2 \beta_1 \beta_2 \mid \: d^2 \alpha_1 \: d^2 \alpha_2 \:
	d^2 \beta_1 \: d^2 \beta_2}, \\
	A_2 \: &\le& \: {\mid \int P(\alpha_1,\alpha_2,\beta_1,\beta_2)
	\alpha_1^* \beta_1^* \alpha_2 \beta_2 \: d^2 \alpha_1 \: d^2
	\alpha_2 \: d^2 \beta_1 \: d^2 \beta_2 \mid \over 2 \int P
	(\alpha_1, \alpha_2, \beta_1, \beta_2) 4 \mid \alpha_1
	\alpha_2 \beta_1 \beta_2 \mid \: d^2 \alpha_1 \: d^2 \alpha_2 \:
	d^2 \beta_1 \: d^2 \beta_2},
\end{eqnarray}
\end{mathletters}
and finally
\begin{equation}
	A_k \le {1 \over 2}, \quad (k=1,2).
\end{equation}

\section{Splitting of an arbitrary single-mode state}
\label{sec:gensplit}

\begin{figure}
\centerline{\psfig{figure=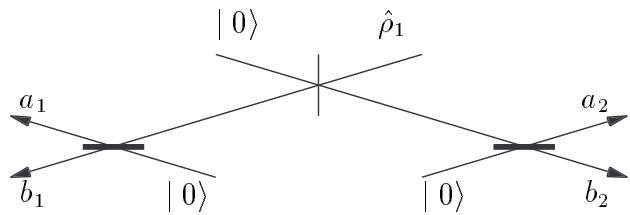}}
\caption{Splitting of an arbitrary single mode state to produce an
EPR state}
\label{fig:gensplit}
\end{figure}

Consider an arbitrary single mode state, defined in terms of the
Sudarshan $P$-distribution as
\begin{equation}
\hat{\rho}_1 = \int P(\alpha) \mid \alpha \rangle \langle \alpha \mid
	d^2 \alpha.
\end{equation}
Assume that this state is mixed with vacuum on a semireflecting
beamsplitter, the total density operator being $\hat{\rho}_2 =
\hat{\rho}_1 \otimes \mid 0 \rangle \langle 0 \mid$ (cf. Fig.
\ref{fig:gensplit}). A coherent state is transformed according to
\begin{equation}
	\mid \alpha \rangle \: \otimes \mid 0 \rangle = \mid {\alpha
	\over \sqrt{2}} \rangle \: \otimes \mid {\alpha \over \sqrt{2}}
	\rangle,
\end{equation}
where the right side is expressed in terms of output states. Thus the
density operator $\hat{\rho}_2$ can be expressed in terms of
the output states as
\begin{equation}
	\hat{\rho}_2 = \int P(\alpha) \mid {\alpha \over \sqrt{2}}
	\rangle \langle {\alpha \over \sqrt{2}} \mid \otimes \mid {\alpha
	\over \sqrt{2}} \rangle \langle {\alpha \over \sqrt{2}} \mid
	d^2 \alpha.
\end{equation}
Next, each output channel is again mixed with vacuum on a
semireflecting beamsplitter (cf. Fig. \ref{fig:gensplit}), the total
density operator now becoming
\begin{equation}
	\hat{\rho}_4 = \rho_2 \: \otimes \mid 0 \rangle \langle 0
	\mid \otimes \mid 0 \rangle \langle 0 \mid.
\end{equation}
In terms of the output states, the density operator then may be
written as
\begin{equation}
	\hat{\rho}_4 = \int P(\alpha) \mid {\alpha \over 2} \rangle_{a_1}
	\: _{a_1}\langle {\alpha \over 2} \mid \otimes \mid {\alpha \over
	2} \rangle_{b_1} \: _{b_1}\langle {\alpha \over 2} \mid \otimes
	\mid {\alpha \over 2} \rangle_{a_2} \: _{a_2}\langle {\alpha
	\over 2} \mid \otimes \mid {\alpha \over 2} \rangle_{b_2} \:
	_{b_2}\langle {\alpha \over 2} \mid d^2 \alpha,
\end{equation}
or, by a change of integration variables,
\begin{equation}
	\hat{\rho}_4 = 4 \int P(2 \beta) \mid \beta \rangle_{a_1} \:
	_{a_1}\langle \beta \mid \otimes \mid \beta \rangle_{b_1} \:
	_{b_1}{\langle} \beta \mid \otimes \mid \beta \rangle_{a_2} \:
	_{a_2}\langle \beta \mid \otimes \mid \beta \rangle_{b_2} \:
	_{b_2}\langle \beta \mid d^2 \beta.
\end{equation}
We now may find the numerator and denominator for the amplitudes
$A_k$ in Eqs. (\ref{eq:amplitudes}). We thus find that
\begin{eqnarray}
	&\langle& ( \hat{a}_1^{\dag} \hat{a}_1 + \hat{b}_1^{\dag}
	\hat{b}_1 ) ( \hat{a}_2^{\dag} \hat{a}_2 + \hat{b}_2^{\dag}
	\hat{b}_2 ) \rangle = \mathrm{Tr} \: \left [ \rho_4 \: (
	\hat{a}_1^{\dag} \hat{a}_1 + \hat{b}_1^{\dag} \hat{b}_1
	) ( \hat{a}_2^{\dag} \hat{a}_2 + \hat{b}_2^{\dag} \hat{b}_2 )
	\right ] \nonumber \\ &=& 16 \: \mathrm{Tr} \: [ \int P(2 \beta)
	\mid \beta \mid^4 \: \mid \beta \rangle_{a_1} \: _{a_1}\langle
	\beta \mid \otimes \mid \beta \rangle_{b_1} \: _{b_1}{\langle}
	\beta \mid \otimes \mid \beta \rangle_{a_2} \: _{a_2}\langle
	\beta \mid \otimes \mid \beta \rangle_{b_2} \: _{b_2}\langle
	\beta \mid d^2 \beta \: ].
\end{eqnarray}
Likewise, we find that
\begin{eqnarray}
	&2& \mid \langle \hat{a}_1^{\dag} \hat{b}_1 \hat{a}_2^{\dag}
	\hat{b}_2 \rangle \mid = 2 \mid \mathrm{Tr} \: \left [ \rho_4 \:
	\hat{a}_1^{\dag} \hat{b}_1 \hat{a}_2^{\dag} \hat{b}_2 \right ]
	\mid \nonumber \\ &=& 8 \: \mathrm{Tr} \: [ \int P(2
	\beta) \mid \beta \mid^4 \: \mid \beta \rangle_{a_1} \:
	_{a_1}\langle \beta \mid \otimes \mid \beta \rangle_{b_1} \:
	_{b_1}{\langle} \beta \mid \otimes \mid \beta \rangle_{a_2} \:
	_{a_2}\langle \beta \mid \otimes \mid \beta \rangle_{b_2} \:
	_{b_2}\langle \beta \mid d^2 \beta \: ]
\end{eqnarray}
and
\begin{eqnarray}
	&2& \mid \langle \hat{a}_1^{\dag} \hat{b}_1^{\dag} \hat{a}_2
	\hat{b}_2 \rangle \mid = 2 \mid \mathrm{Tr} \: \left [ \rho_4 \:
	\hat{a}_1^{\dag} \hat{b}_1^{\dag} \hat{a}_2 \hat{b}_2 \right ]
	\mid \nonumber \\ &=& 8 \: \mathrm{Tr} \: [ \int P(2
	\beta) \mid \beta \mid^4 \: \mid \beta \rangle_{a_1} \:
	_{a_1}\langle \beta \mid \otimes \mid \beta \rangle_{b_1} \:
	_{b_1}{\langle} \beta \mid \otimes \mid \beta \rangle_{a_2} \:
	_{a_2}\langle \beta \mid \otimes \mid \beta \rangle_{b_2} \:
	_{b_2}\langle \beta \mid d^2 \beta \: ].
\end{eqnarray}
It thus follows that
\begin{equation}
	\langle ( \hat{a}_1^{\dag} \hat{a}_1 + \hat{b}_1^{\dag}
	\hat{b}_1 ) ( \hat{a}_2^{\dag} \hat{a}_2 + \hat{b}_2^{\dag}
	\hat{b}_2 ) \rangle = 4 \mid \langle \hat{a}_1^{\dag} \hat{b}_1
	\hat{a}_2^{\dag} \hat{b}_2 \rangle \mid = 4 \mid \langle
	\hat{a}_1^{\dag} \hat{b}_1^{\dag} \hat{a}_2 \hat{b}_2 \rangle
	\mid,
\end{equation}
and, by substituting into Eqs. (\ref{eq:amplitudes}), we conclude
that
\begin{equation}
	A_k = \frac{1}{2} \quad (k=1,2).
\end{equation}

\begin{multicols}{2}

\end{multicols}

\end{document}